\documentclass[]{pasj01}

\newenvironment{contribution}{\section*{Author contributions}\fontsize{8}{11}\selectfont}{\par}

\begin{document} 
\Received{2017/07/21}
\Accepted{2017/07/27}

\title{Hitomi X-ray studies of Giant Radio Pulses from the Crab pulsar
\thanks{Corresponding authors are
Yukikatsu \textsc{Terada},
Teruaki \textsc{Enoto},
Shu \textsc{Koyama},
Aya \textsc{Bamba},
Toshio \textsc{Terasawa},
Shinya \textsc{Nakashima}, 
Tahir \textsc{Yaqoob},
Hiromitsu \textsc{Takahashi},
and
Shin \textsc{Watanabe}.
}
}
\author{Hitomi Collaboration,
Felix \textsc{Aharonian}\altaffilmark{1},
Hiroki \textsc{Akamatsu}\altaffilmark{2},
Fumie \textsc{Akimoto}\altaffilmark{3},
Steven W. \textsc{Allen}\altaffilmark{4,5,6},
Lorella \textsc{Angelini}\altaffilmark{7},
Marc \textsc{Audard}\altaffilmark{8},
Hisamitsu \textsc{Awaki}\altaffilmark{9},
Magnus \textsc{Axelsson}\altaffilmark{10},
Aya \textsc{Bamba}\altaffilmark{11,12},
Marshall W. \textsc{Bautz}\altaffilmark{13}
% ...
}
\altaffiltext{1}{Dublin Institute for Advanced Studies, 31 Fitzwilliam Place, Dublin 2, Ireland}
\altaffiltext{2}{SRON Netherlands Institute for Space Research, Sorbonnelaan 2, 3584 CA Utrecht, The Netherlands}
\altaffiltext{3}{Institute for Space-Earth Environmental Research, Nagoya University, Furo-cho, Chikusa-ku, Nagoya, Aichi 464-8601}
\altaffiltext{4}{Kavli Institute for Particle Astrophysics and Cosmology, Stanford University, 452 Lomita Mall, Stanford, CA 94305, USA}
\altaffiltext{5}{Department of Physics, Stanford University, 382 Via Pueblo Mall, Stanford, CA 94305, USA}
\altaffiltext{6}{SLAC National Accelerator Laboratory, 2575 Sand Hill Road, Menlo Park, CA 94025, USA}
\altaffiltext{7}{NASA, Goddard Space Flight Center, 8800 Greenbelt Road, Greenbelt, MD 20771, USA}
\altaffiltext{8}{Department of Astronomy, University of Geneva, ch. d'\'Ecogia 16, CH-1290 Versoix, Switzerland}
\altaffiltext{9}{Department of Physics, Ehime University, Bunkyo-cho, Matsuyama, Ehime 790-8577}
\altaffiltext{10}{Department of Physics and Oskar Klein Center, Stockholm University, 106 91 Stockholm, Sweden}
\altaffiltext{11}{Department of Physics, The University of Tokyo, 7-3-1 Hongo, Bunkyo-ku, Tokyo 113-0033}
\altaffiltext{12}{Research Center for the Early Universe, School of Science, The University of Tokyo, 7-3-1 Hongo, Bunkyo-ku, Tokyo 113-0033}
\altaffiltext{13}{Kavli Institute for Astrophysics and Space Research, Massachusetts Institute of Technology, 77 Massachusetts Avenue, Cambridge, MA 02139, USA}
% ...
\email{terada@phy.saitama-u.ac.jp}


\maketitle

\begin{abstract}
To search for giant X-ray pulses correlated with the giant radio pulses (GRPs) from the Crab pulsar, we performed a simultaneous observation of the Crab pulsar with the X-ray satellite Hitomi in the 2 -- 300 keV band and the Kashima NICT radio observatory in the 1.4 -- 1.7 GHz band with a net exposure of about 2 ks on 25 March 2016, just before the loss of the Hitomi mission.
The timing performance of the Hitomi instruments was confirmed to meet the timing requirement and about 1,000 and 100 GRPs were simultaneously observed at the main and inter-pulse phases, respectively, and we found no apparent correlation between the giant radio pulses and the X-ray emission in either the main or inter-pulse phases.
All variations are within the 2 sigma fluctuations of the X-ray fluxes at the pulse peaks, and the 3 sigma upper limits of variations of main- or inter- pulse GRPs are 22\% or 80\% of the peak flux in a 0.20 phase width, respectively, in the 2 -- 300 keV band.
The values become 25\% or 110\% for main or inter-pulse GRPs, respectively, 
when the phase width is restricted into the 0.03 phase.
Among the upper limits from the Hitomi satellite, those in the 4.5-10 keV and the 70-300 keV are obtained for the first time, and those in other bands are consistent with previous reports.
Numerically, the upper limits of main- and inter-pulse GRPs in the 0.20 phase width are about (2.4 and 9.3) $\times 10^{-11}$ erg cm$^{-2}$, respectively. 
No significant variability in pulse profiles implies that the GRPs originated from a local place within the magnetosphere and the number of photon-emitting particles temporally increases.
However, the results do not statistically rule out variations correlated with the GRPs, because the possible X-ray enhancement may appear due to a $>0.02$\% brightening of the pulse-peak flux under such conditions. 
\end{abstract}

\section{Introduction}
\label{section:introduction}

Giant Radio Pulses (GRPs) consist of sporadic and short-lived radiation,
	during which time the radio flux density becomes 2--3 orders of magnitudes 
	brighter than the regular, averaged pulse flux density. 
So far, this phenomenon has been discovered 
	in $\sim$14 radio pulsars 
	(for a review, see \cite{2006ChJAS...6b..41K} and references therein),
	including both ``traditional'' rotation-powered pulsars 
	(e.g., the Crab pulsar)
	and millisecond pulsars (e.g., PSR~B1937+21).
Although the emission mechanism of the GRPs is still unknown,
	previous radio studies have shown some distinctive properties of the GRPs.
The typical temporal width of individual GRPs 
	is narrow,
	spanning a range from a few nanoseconds to a few microseconds \citep{2003Natur.422..141H}.
GRPs occur in certain pulse phases with no clear periodicity. 
The energy spectrum of GRPs follows
	a power-law distribution \citep{2007A&A...470.1003P,2016ApJ...832..212M}, 
	different from the Gaussian or log-normal distribution of the normal pulses
	\citep{2012MNRAS.423.1351B}.
Since studies of the ordinary pulses can only provide average information from
	the pulsar magnetosphere,
	observations of GRPs are imperative for furthering our understanding of the pulsar radiation mechanism. 
More recently, a 
	hypothetical proposal of GRPs from young pulsars 
	as candidates for the origin of fast radio bursts (FRBs; \cite{2016MNRAS.457..232C})
        have been attracting more and more attention.
	These phenomena are  extragalactic bright radio transients with $\sim$1~msec duration 
	\citep{2007Sci...318..777L,2013Sci...341...53T,2017Natur.541...58C}.
Seeking to reveal properties of known GRPs, such as identifying their counterparts in other wavelengths, 
	is also a key line of investigation to examine the young pulsar model that could explain FRBs 
	(e.g., \cite{2016MNRAS.460.2875Y, 2016ApJ...832L...1D}). 

The Crab pulsar (PSR~B0531+21)
	is one of the most intensively studied rotation-powered pulsars 
	since the initial discovery of its GRPs \citep{1968Sci...162.1481S}. 
This famous pulsar exhibits GRPs
	occurring both in the main pulse and the interpulse,
	which have been mainly studied at radio wavelengths	
	\citep{2007A&A...470.1003P}.	
Since the pulsed energy spectrum of the Crab pulsar 
	covers a wide range, from the coherent radio emission 
	to the incoherent high-energy radiation 
	at optical, X-rays, and gamma-rays,
	there have been multi-wavelength campaigns to attempt
	to search for enhancements 
	at higher energy bands, simultaneous with the GRPs.
In the optical band,
	a significant 3\% optical enhancement was discovered 
	with 7.2$\sigma$ significance 
	from the main pulse peak phase 
	by the Westerbork Synthesis Radio Telescope 
	and the 4.2-m William Herschel Telescope
	 \citep{2003Sci...301..493S}.
This result was further confirmed 
	by the Green Bank Telescope and 
	the Hale telescope \citep{2013ApJ...779L..12S}. 
These detections imply that the coherent radio emission is somehow linked to 
	the incoherent higher energy (optical) radiation. 
Despite intensive efforts to search at even higher energy bands,
	so far there are only upper-limits in soft X-rays and higher energy bands.
	Reports of these upper limits can be found for
	soft X-rays (Chandra, 1.5--4.5\,keV; \cite{2012ApJ...749...24B}), 
	for soft gamma-rays (CGRO/OSSE, 50--220\,keV; \cite{1995ApJ...453..433L}), 
	for gamma-rays (Fermi/LAT, 0.1--5\,GeV; \cite{2011ApJ...728..110B}), 
	and for very high energy gamma-rays (VERITAS, $>$150\, GeV; \cite{2012ApJ...760..136A}). 

The X-ray astronomical satellite Hitomi (ASTRO-H) 
	was launched on February 17, 2016
	via a H-IIA launch vehicle
	from Tanegashima Space Center in Japan,
	and successfully entered into a low Earth orbit
	at an altitude of 575\,km \citep{ASTROH2016}.
The satellite is designed to cover a wide energy range from 0.3\,keV up to 600\,keV
	with four new X-ray instruments: 
	the microcalorimeter (Soft X-ray Spectrometer, SXS; \cite{ASTROH_SXS}),
	a wide field-of-view X-ray CCD detector (Soft X-ray Imager, SXI; \cite{ASTROH_SXI}),
	two Si/CdTe hybrid hard X-ray imagers (Hard X-ray Imager, HXI; \cite{ASTROH_HXI}),
	and a Compton telescope (Soft Gamma-ray Detector, SGD; \cite{ASTROH_SGD}).
Such a wide energy coverage with high time resolution at a few microseconds \citep{2017Hitomi_Time_system} made Hitomi
	suitable for the search for X-ray enhancement simultaneously with the GRPs. 
After initial operations before opening the gate valve of the SXS, 
        including successful observations of the Perseus cluster of galaxies 
        \citep{2016Natur.535..117H} and other supernova remnants
        (e.g., N132D and G21.5-0.9) 
        the spacecraft lost communications with the ground stations on March 26, 
        and eventually the mission was terminated.
Therefore, the energy coverage below 2~keV was lost for the SXS.

On March 25, just before the satellite loss,
	we observed the Crab pulsar with Hitomi 
	for onboard instrumental calibration activity. 
The requirement and goal of the absolute timing accuracy of the Hitomi satellite
	are 350\,$\mu$s and 35\,$\mu$s, respectively \citep{2017Hitomi_Time_system}.
In order to verify the timing tag accuracy,
	we compared arrival times of  
	the main pulse peak of the Crab pulsar
	with the radio or X-ray ephemeris provided by other observatories
	\citep{2008PASJ...60S..25T}.
The archival monthly radio ephemeris of the Crab pulsar has been regulary provided 	
	by the Jodrell-Bank observatory\footnote{http://www.jb.man.ac.uk/~pulsar/crab/crab2.txt}
	on every 15th monitoring 
	 \citep{1993MNRAS.265.1003L}.
Interpolating from this,
	the predicted ephemeris of the Crab pulsar timing 
        in Barycentric Dynamical Time (TDB)
        is tabulated in table~\ref{table:ephemeris}.


Arrival times of the radio pulses are known to be delayed from X-rays 
	in proportion to the interstellar dispersion measure (DM). 
The long-term, averaged DM of the Crab pulsar 
	is $\sim$56.8\,pc\,cm$^{-3}$,
	corresponding to $\sim$120\,ms delay 
	of the 1.4\,GHz radio pulses relative to X-rays.
Although this radio delay is corrected via radio analyses (de-dispersion),
	this DM is known to show fluctuation in time
	with $\sim$0.028\,pc\,cm$^{-3}$ (1$\sigma$) of a Gaussian distribution.
This corresponds to an intrinsic uncertainty of $\sim$60\,$\mu$s timing accuracy, 
	higher than our goal for the timing accuracy (35\,$\mu$s). 
Therefore, 	
	we coordinated follow-up radio observations 
	simultaneous with our X-ray observations
	to reduce uncertainties due to this fluctuating DM.
In this paper, 	
	we report X-ray studies 
	of GRPs from the Crab pulsar 
	based on simultaneous X-ray and radio observations.
Detailed investigations of the instrumental timing calibration will be summarized  
	in a different paper \citep{2017Hitomi_Time_inorbit}.

\begin{table}
  \tbl{Crab ephemeris in Radio}{%
  \begin{tabular}{ll}
      \hline
      Parameter & Value \\
      \hline
      Main pulse & MJD 57472.0000002874260532 \\
      Period & 0.0337204396077250 s\\
      Pediod derivative &  4.1981605$\times 10^{-13}$ s s$^{-1}$\\
      \hline
    \end{tabular}}\label{table:ephemeris}
\begin{tabnote}
Ephemeris of the Crab pulsar determined by the radio observations
on 25 March 2016. The 'Main pulse' represents the arrival time of 
the main pulse in the radio band. All values are in TDB.
\end{tabnote}
\end{table}

\section{Observation and Data Reduction}
\label{section:observation_dataReduction}

\subsection{X-ray and radio simultaneous observation of the Crab pulsar}
\label{section:observation_dataReduction:observation}
The X-ray observation of the Crab pulsar was made with all the instruments on board the Hitomi satellite, 
	starting from 12:17 on March 25, 2016 until 18:01 (UT) [TDB]
	with a total on-source duration of 9.7\,ks.
The radio observations of the Crab pulsar were made in two frequency bands,
(a) 1.4 -- 1.7GHz at the Kashima observatory  from 03:00:00 to 14:00:00 UTC,
and 
(b) 323.1 -- 327.1MHz at the Iitate observatory from 09:30:00 to 13:00:00 UTC.
The locations of the observatories are listed in table 1 of \citet{2016ApJ...832..212M}. 
After the start time of the X-ray observation
we idenfitied 3350 GRPs (section~\ref{section:GRPselection}) for (a),
but only 94 GRPs for (b).
In terms of the occurence probability (number of GRPs per minute),
the ratio between (a) and (b) was $\sim$19:1.
\citet{2016ApJ...832..212M} reported, on the other hand,
that the ratio was $\sim$3:1 on 6 -- 7 September 2014.
The marked difference between these ratios seems to be caused by 
refractive interstellar scintilation (RISS; e.g. Lundgren et al. (1995)):
While the RISS condition for 1.4 -- 1.7GHz would have corresponded 
to a phase of the intensity larger than the average, 
the RISS condition for 325MHz would have corresponded to a phase
of the intensity smaller than the average.
Therefore we concentrate on observation (a) in what follows.

\subsection{Data Reduction of Radio observations}
\label{section:RadioObservation} 
\def\FreqTable{{
\begin{table*}
  \tbl{Frequency bands of the Kashima observatory}{%
  \begin{tabular}{cccc}
      \hline
      channel & frequency (MHz) &  original bandwidth (MHz)  & effective bandwidth (MHz)    \\
       $k$    & minimum-maximum &                            & after filterization ($\Delta \nu_k$)\\
      \hline
      ch0 & 1404-1436 & 32 & 20.48 \\
      ch1 & 1570-1602 & 32 & 16.00 \\
      ch2 & 1596-1628 & 32 & -     \\
      ch3 & 1622-1654 & 32 & -     \\
      ch4 & 1648-1680 & 32 & 24.83 \\
      ch5 & 1674-1706 & 32 & 21.82 \\
      ch6 & 1700-1732 & 32 & 23.23 \\
      (ch7) & (1702-1734) & (32) & -     \\
      \hline
    \end{tabular}}
\label{table:FreqTable}
\begin{tabnote}
The frequency range between ch0 and ch1, 1436-1570MHz, is avoided
so as to minimize the radio frequency interference (RFI)
from cellphone base stations.
Only the right-hand circular polarized signals were received.
In the coherent dedispersion process for each channel,
the timing of the voltage data is adjusted
to that for the maximum frequency of the channel.
\end{tabnote}
\end{table*}
}}

\def\FreqTable{{
\begin{table*}
  \tbl{Frequency bands of the Kashima observatory}{%
  \begin{tabular}{cccc}
      \hline
      channel & frequency (MHz) &  original bandwidth (MHz)  & effective bandwidth (MHz)    \\
       $k$    & minimum-maximum &                            & after filterization ($\Delta \nu_k$)\\
      \hline
      ch0 & 1404-1436 & 32 & 20.48 \\
      ch1 & 1570-1602 & 32 & 16.00 \\
      ch2 & 1596-1628 & 32 & -     \\
      ch3 & 1622-1654 & 32 & -     \\
      ch4 & 1648-1680 & 32 & 24.83 \\
      ch5 & 1674-1706 & 32 & 21.82 \\
      ch6 & 1700-1732 & 32 & 23.23 \\
      (ch7) & (1702-1734) & (32) & -     \\
      \hline
    \end{tabular}}
\label{table:FreqTable}
\begin{tabnote}
The frequency range between ch0 and ch1, 1436-1570MHz, is avoided
so as to minimize the radio frequency interference (RFI)
from cellphone base stations.
Only the right-hand circular polarized signals were received.
In the coherent dedispersion process for each channel,
the timing of the voltage data is adjusted
to that for the maximum frequency of the channel.
\end{tabnote}
\end{table*}
}}

\def\myFigOne{{
\begin{figure*}
 \begin{center}
   \includegraphics[width=1.0\textwidth]{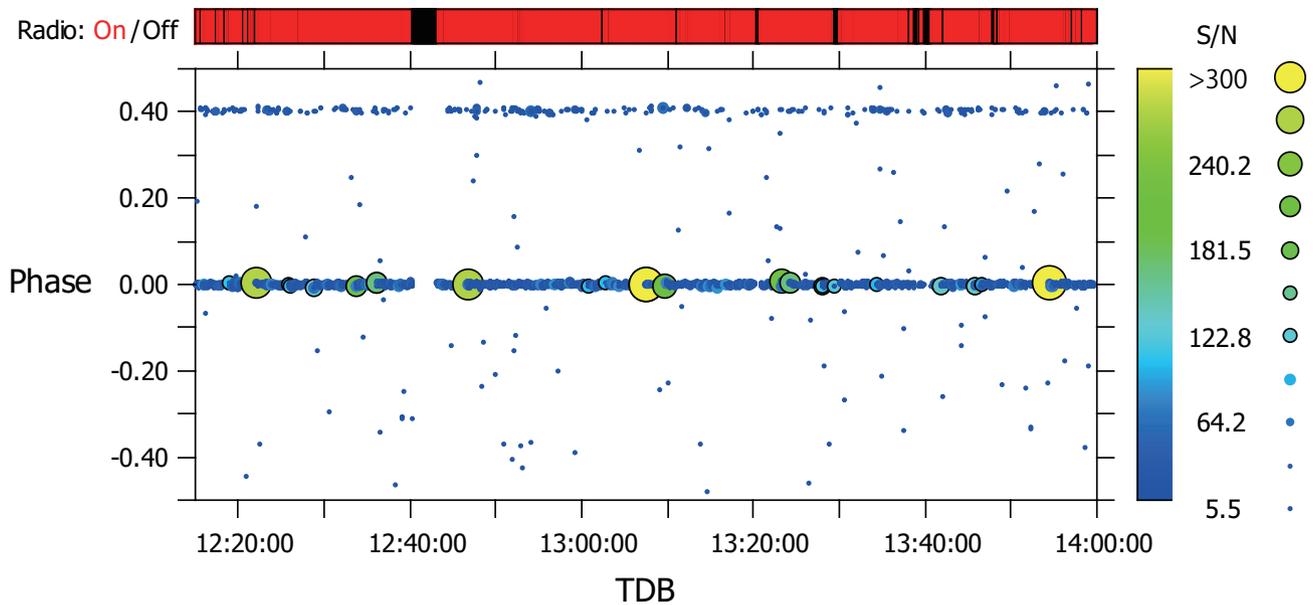}
 \end{center}
\caption{
In the main panel
GRP candidates (S/N$>$5.5) are shown in the (TDB, $\varphi$) plane,
where two clusters in $\varphi$ are of main pulse and interpulse GRPs (see text).
Scatterd points show the remaining noise contribution.
The threshold S/N=5.5 corresponds to the minimum pulse energy 2.2 kJy~$\mu$s.
The strongest main pulse occurred at 13:07:25.645TDB had the peak S/N$\sim$ 659.
It spreaded over $\sim$40$\mu$s interval having the total pulse energy 358 kJy~$\mu$s.
}
\label{fig:GRPcandidateSNgt55}
\end{figure*}
}}

\def\dummy{{
Data reduction radio (TBA).
\textcolor{blue}{Table \ref{table:ephemeris} is prepared here.}

\textcolor{blue}{comment(YT-20170419): Section 2.2 and 2.3 were swapped.}
- how the radio data were cleaned up?
- how the GRPs are identified?
- statistics of GRPs of main pulse / inter pulse.
- \textcolor{blue}{(A/I) ask Terasawa-san et al to fill this section.}
}}

\subsubsection{Frequency assignment}
The radio observation in the 1.4 -- 1.7GHz band was made with the 34m telescope
at the Kashima Space Technology Center \citep{2016PASP..128h4502T}
operated by the NICT (National Institute of Information and Communications Technology).
We used the ADS3000+ recorder \citep{2010ivs..conf..378T}
which has a capability of 8 individual channels
with 4-bit 64MHz Nyquist-rate sampling.
(The sampling time step $\delta t$ is 1/64 MHz=15.625ns,
and the data rate 2Gbit/s.)
table~\ref{table:FreqTable} shows the frequency assignment for 8 channels.
Channel 7, the backup for channel 6 with a slight frequency shift, was not used for the following data analysis.

\FreqTable

\subsubsection{Determination of DM}
\label{section:DeterminationDM}
We first determined the dispersion measure (DM) appropriate for
the epoch of the observation, 25 March 2016 (MJD 57472).
While the Jodrell Bank Crab pulsar monthly ephemeris
reports the values of 56.7657 pc cm$^{-3}$ (=DM$_{\rm JB}$) for 15 March 2016 (MJD 57462),
and 56.7685 pc cm$^{-3}$ for 15 April 2016 (MJD 57493),
we should take into account possible intra-month variations of DM 
which are sometimes very erratic (e.g., \cite{2008AA...483...13}).
With DM$_{\rm JB}$ as a trial value, we coherently dedispersed 
\citep{1975MCP14,2004HPA} the ch0 data and found several bright GRPs.
We then extended the dedispersion analysis to all the channels
for $\sim$50 ms intervals including these GRPs.
The best value of DM, 56.7731$\pm 0.0001$  pc cm$^{-3}$ (=DM$_{\rm best}$),
was obtained so as to get the alignment of the substructures of these GRPs
(e.g., \cite{1999ApJ...517...460})
in all channels with $\sim0.1\mu$s accuracy.
An example of a successful alignment can be seen in Figure 2 of 
\citet{2016ApJ...832..212M}.
The frequency bands LL and LH approximately correspond to ch1 and ch4-6 here.

\subsubsection{Frequency-domain RFI rejection}
\label{section:FrequencyDomainRFIrejection}
During the process of finding DM$_{\rm best}$, 
we noticed that two channels, ch2 and ch3, were severly contaminated by 
radio frequency interferences (RFI).
Since RFI occurred intermittently, we could still use these channels
for the bright GRP search.
However, for weaker GRPs search,
we excluded ch2 and ch3 from the following analysis.
We further noticed that other channels 
were weakly contaminated by RFI in some limited frequency ranges.
To minimize the effect of RFI, we filtered out these
contaminated frequency ranges.
The numerical filter was applied
at the first stage of the coherent dedispersion process,
where the time series of antenna voltage data
are subjected to FFT (fast Fourier transformation)
and decomposed into Fourier components.
For the RFI-contaminated frequency ranges, we set their Fourier components to zero.
The overlapping frequency ranges for ch4-ch5 and ch5-ch6 are also filtered out at this stage.
The rightmost column of table~\ref{table:FreqTable}
gives the resultant effective bandwidths after filtering.
The total effective bandwidth, $\Delta \nu_{\rm sum}  =\Delta \nu_0 + \Delta \nu_1 + ...+ \Delta \nu_6$,
is 106.36MHz.

\myFigOne

\subsubsection{GRP selection}
\label{section:GRPselection}
In the main panel of figure \ref{fig:GRPcandidateSNgt55}
dots show all GRP candidates with a S/N (signal to noise ratio) $>$5.5
(or pulse energy $>\sim$2.2 kJy~$\mu$s, 
see Appendix~\ref{section:appendix_radio}),
where dots are sized and color-coded with the values of S/N. 
The abscissa and ordinate of the panel represent 
the time in TDB and the pulsar rotation phase $\varphi$ respectively. 
The top panel of figure \ref{fig:GRPcandidateSNgt55}
shows the time intervals subjected to
the time-domain RFI rejection (see Appendix~\ref{section:appendix_radio}) 
in black (OFF), and the time intervals kept, in red (ON).
In total, 571s (5729s) in time intervals were rejected (kept).

As can be seen in figure \ref{fig:GRPcandidateSNgt55}
there are two clusters of GRP candidates in $\varphi$.
We adjusted the initial value of $\varphi$ at 00:00:00TDB
($y_0$ in (\ref{eqn:getPhi})) so as to locate the peak of the main cluster
at $\varphi$=0, which is the main pulse GRPs 
(hereafter we call them the ``MP-GRP'').
The second cluster found around $\varphi=$0.4056
corresponds to the inter-pulse GRPs (hereafter ``IP-GRP'').
Scattered points also seen in figure \ref{fig:GRPcandidateSNgt55}
are due to the noise component.
With the selection criteria,
(1) $-0.0167 \le \varphi \le +0.0167$ for the MP-GRPs
and
(2) $0.3889 \le \varphi \le 0.4222$ for the IP-GRPs,
we identified
3090 MP-GRPs and 260 IP-GRPs
during the interval between 12:15:00TDB and 14:00:00TDB.
We estimated the noise contributions in terms of fake GRPs
to be $11 \pm 3$ ($0.4 \pm 0.1$\% and $4 \pm 1$\% 
for the MP- and IP-GRPs, respectively.)
The pulse energy distributions of GRPs have power-law shapes
with the spectral indices $-2.88 \pm 0.52$ for the MP-GRPs
and $-2.91 \pm 1.13$ for the IP-GRPs.


\subsection{Data Reduction of Hitomi observation}
\label{section:observation_dataReduction:data_xray}

The X-ray data obtained with the Hitomi satellite were processed by the standard Hitomi pipeline 
version 03.01.005.005 (Angelini et al 2016) with the pre-pipeline version 003.012.004p2.004 
using the hitomi ftools in the HEAsoft version 6.20, with 
CALDB versions gen20161122, hxi20161122, sgd20160614, sxi20161122, and sxs20161122.
In the timing analyses of Hitomi data in the following sections, 
the SXI and SGD-2 data were not used, 
because the timing resolution of the SXI was insufficient for the analyses and 
the SGD-2 was not in the nominal operation mode during the simultaneous epoch 
with the radio observation. 

\begin{figure}
 \begin{center}
  \includegraphics[width=0.50\textwidth,angle=0]{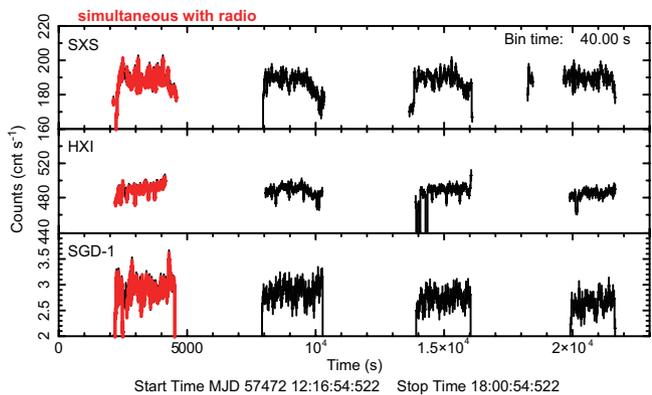} 
 \end{center}
\caption{Light curve of the Crab with from Hitomi SXS, HXI within region a, HXI within region b, and SGD-1, from top to bottom panels, respectively. The black croses represent the entire clened events of the Crab with Hitomi, and red shows the same but within the simultaneous intervals with radio observatories.}\label{fig:lightcurve}
\end{figure}

The standard cleaned events were used for the SXS and HXI analyses; 
the low resolution events (ITYPE == 3 or 4; \cite{ASTROH_SXS}) of the SXS events were not excluded 
in order to maximize the statistics, 
although the time resolution of the low resolution events was worse (80 $\mu$s) 
than those of high or med resolution events (5 $\mu$s). 
The HXI data were extracted using a sky image region around the Crab pulsar, 
out to 70 arcsec radius from the image centroid.
On the analyses of the SGD-1 data, the photo-absorption events were extracted 
as described in Appendix~\ref{section:appendix:sgd}.
At this stage, the total exposure times of the Hitomi Crab observation were
9.7, 8.0, and 8.6 ksec for the SXS, HXI, and SGD, respectively.
The background-inclusive light curves of these data were shown in Fig.\  \ref{fig:lightcurve} black.
Note that no energy selection were applied to the events; the rough energy band for the SXS, HXI, and SGD-1 photo absorption events were 2 -- 10 keV, 2 -- 80 keV, and 10 -- 300 keV, respectively.

The TIME columns of all the event lists of SXS, HXI, and SGD-1 were converted 
into a barycentric position using the ``barycor''  ftool in the hitomi package 
of HEAsoft 6.20 and the hitomi orbital file \citep{2017Hitomi_Time_system}. 
The target position for the barycentric correction was 
(R.A., DEC) = (\timeform{83D.633218}, \timeform{+22D.014464}) for this analyses.
The period and period derivatives determined only with the Hitomi data were consistent with the ephemeris 
from the radio summarized in table~\ref{table:ephemeris}.
As described in \citet{2017Hitomi_Time_system} and \citet{2017Hitomi_Time_inorbit},
the time differences between instruments 
were negligible for the timing analyses of the giant radio pulses reported here.

Finally, all the good time intervals of the radio observation were applied to the 
Hitomi Crab data, which then results in a shorter duration, as shown in Fig.\  \ref{fig:lightcurve} (red). 
Consequently, the total exposure times for the SXS, HXI, and SGD-1 that were simultaneously observed 
with the radio observatories become 2.7, 1.7, and 2.1 ks, respectively.
About $10^3$ GRP cycles were exposed among $(5$ -- $6)\times 10^4$ cycles 
by each instrument, as summarized in table~\ref{table:grp_stat}.

\begin{table}
  \tbl{Statistics of GRPs during the simultaneous observation}{%
  \begin{tabular}{ccccc}
      \hline
      Instrument & Exposure & \# of cycles 
    & MP-GRPs & IP-GRPs \\
      \hline
      SXS & 2.2 ks & 64,701 & 1,171 & 103 \\
      HXI & 1.7 ks & 50,705 & 945 & 85 \\
      SGD & 2.1 ks & 63,197 & 1,144 & 98 \\
      \hline
    \end{tabular}}\label{table:grp_stat}
\begin{tabnote}
Total exposures, number of cycles of pulsar pulses,
and number of GRPs for main and inter pulses,
during the simultaneous observation 
between radio observatories and Hitomi instruments.
\end{tabnote}
\end{table}

\section{Analyses and Results}
\label{section:analyses_results}

In this section, we used the cleaned events of Hitomi SXS/HXI/SGD instruments
obtained at the end of the section \ref{section:observation_dataReduction:data_xray} 
and the pulsar ephemeris in table~\ref{table:ephemeris}.

\subsection{Variation of the Pulse Profiles in GRPs}
\label{section:analyses_results:pulse_enhancement}

\begin{figure*}
 \begin{center}
  \includegraphics[width=0.75\textwidth,angle=0]{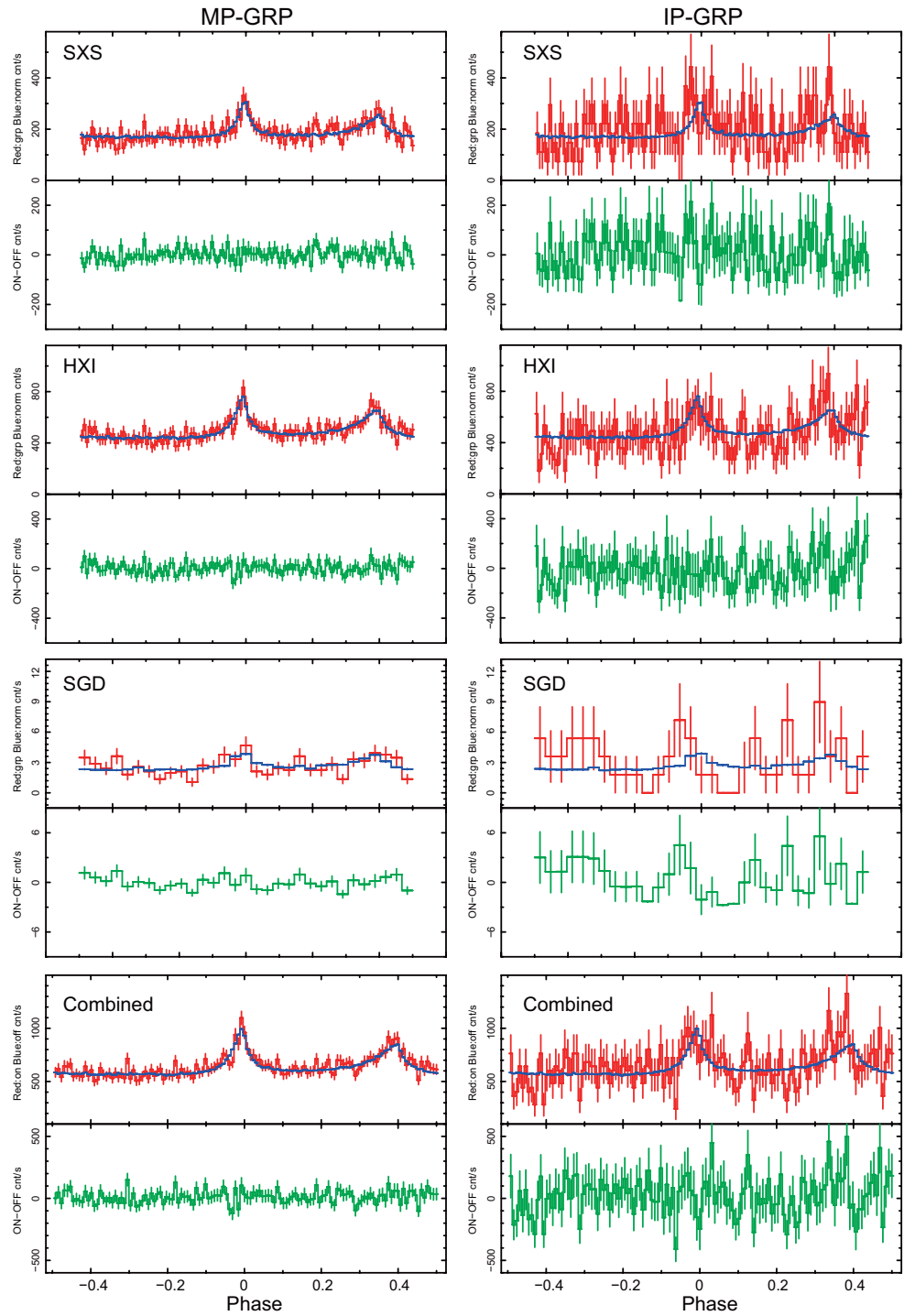}
 \end{center}
\caption{Comparison of Crab pulse profiles between the NORMAL and the GRP cycles, 
which were shown in blue and red, respectively, 
and the green croses in the each bottom panel show the difference between them.
The left and right panels were the plots of the MP-GRPs and IP-GRPs, respectively,
and from the top to the bottom, the data taken by the SXS, HXI, SGD-1, and combined data
were plotted, respectively.}
\label{fig:ana_1pulse}
\end{figure*}

Most significant MP- and IP-GRPs, shown as large circles in figure~\ref{fig:GRPcandidateSNgt55}, were detected at 12:46:44 and 12:54:10 (TDB) on 25 May 2016, respectively, but no significant variations were seen in the X-ray photons before and after the GRPs.
Therefore, we then try to stack X-ray events which were correlated with MP- or IP-GRPs to see a possible enhancement in the X-ray band.
X-ray events within one cycle of each MP-GRP (hereafter we call them the ``MP-GRP cycles'') were accumulated between $\varphi=$ -0.5 to +0.5 phases 
from the arrival time of the main pulse of the radio-defined MP-GRPs (i.e., $\varphi=0$). 
The events outside the MP-GRP cycles were defined as ``NORMAL cycles''
and were accumulated for comparison.
Both groups of events were folded by the radio ephemeris in table~\ref {table:ephemeris} 
to see the pulse profiles of MP-GRP and NORMAL cycles. 
As shown in the top panels of the left-hand plots in figure \ref{fig:ana_1pulse}, 
no major enhancements could be seen between the two profiles.
The difference between the two, shown in the bottom panels, was consistent with being statistically constant 
among all the instruments and along all of the phases ($-0.5 \leq \varphi \leq 0.5$).
Note that the pulse profile of the Crab pulsar with the SXS is free from a possible distortion by the dead time, which occurs in $> 5$ s on the SXS.
Similarly, the distortion of the profiles of the HXI and SGD can be also ignored in comparison between the GRP and NORMAL shapes, although the absolute fractions of the dead to live times were about 75\% (Section \ref{section:analyses_results:upper_limit_flux}).
The same analyses were performed for the inter-pulse GRPs (hereafter, ``IP-GRP cycles''),
and no significant enhancement between pulse profiles at IP-GRP and another NORMAL cycles was found, 
as seen in figure \ref{fig:ana_1pulse} (right).
The statistical errors were very high on the Hitomi datasets, both in MP- and IP-GRPs.

\begin{figure}
 \begin{center}
  \includegraphics[width=0.35\textwidth,angle=0]{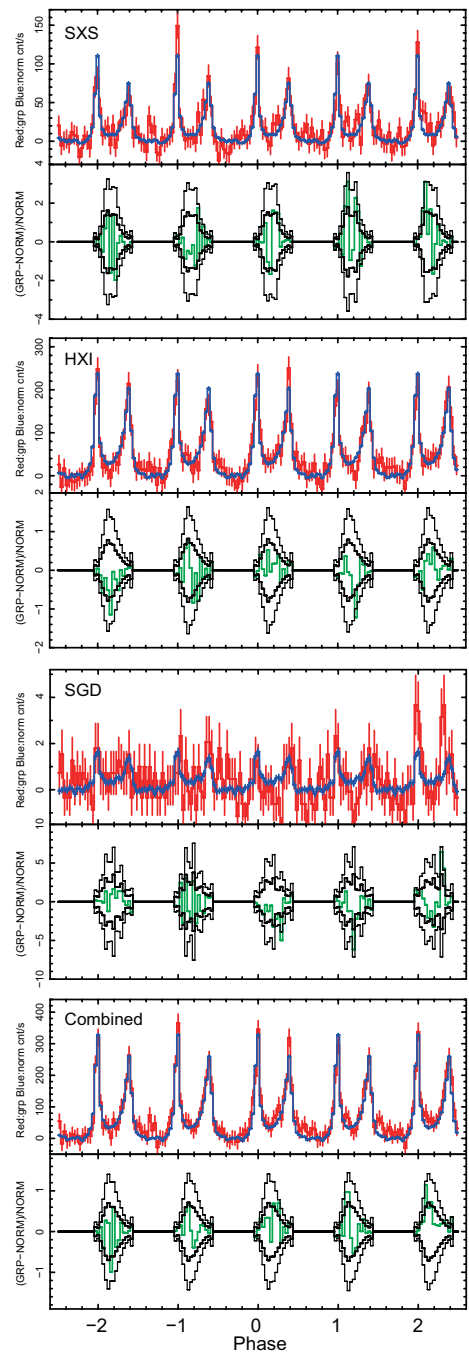}
 \end{center}
\caption{The top plots in each panel show the same pulse profiles of NORMAL and on/near GRP cycles represented by blue and red lines, respectively.
The counts in off-phase ($\varphi = $ 0.6 -- 0.8) of the NORMAL cycles were subtracted from these pulse profiles. 
The bottom plots in each panel represent the ratio of the enhancement of near GRP 
data relative to the NORMAL cycles, which was shown in green.
The data during the OFF phase were not plotted here.
The statistical uncertainties of each phase bin at 1 and 2 sigma were shown 
in thick and thin black lines, respectively. 
The SXS, HXI, SGD-1, and combined data were shown from top to bottom panels, respectively.}
\label{fig:ana_5pulses}
\end{figure}

In order to see some possible enhancements in several cycles around the GRPs in a wider time range, 
we then accumulated the events from 2 cycles before, to 2 cycles after the MP-GRPs;
i.e., five pulses $-2.5 < \varphi < 2.5$ were plotted where $-0.5 \leq \varphi \leq 0.5$ corresponds to the MP-GRP cycle. 
Similar to the previous single-pulse analyses, 
the NORMAL cycles, here, were defined outside the 5 cycles around the MP-GRPs. 
The results were shown in figure \ref {fig:ana_5pulses}.
According to the time intervals between GRPs, about 0.7 \% and 2.4\% of MP-GRPs were contaminated within $\pm 1$ or $\pm 2$ cycles from the GRP, respectively.
To estimate the statistical errors on the pure-pulsed components, 
the non-pulsed counts accumulated from the OFF phase ($\varphi=$ 0.6 -- 0.8)
were subtracted from the pulse profiles of the MP-GRP and NORMAL cycles. 
Several possible enhancements could be seen in several main pulses in the soft energy band by the SXS 
in the top panels of figure \ref {fig:ana_5pulses}, 
however, the significance was all below 2 $\sigma$ as indicated in the bottom panels, 
and no corresponding enhancement was seen in the hard X-ray band by the HXI.
The same study could be performed for IP-GRPs but the statistical errors were very high 
and the results were the same as for the MP-GRP cases. 
Therefore, no enhancements were detected in all phases among five cycles around GRPs from the Hitomi data.

\subsection{Pulse peak enhancement at GRPs }
\label{section:analyses_results:pulse_peak_enhancement}

Since no significant enhancement found in five cycles before/after GRPs (section~\ref{section:analyses_results:pulse_enhancement}), we then concentrated on the statistical tests of possible enhancements at the peak of pulses.
Here, we compared the non-pulse subtracted peak-counts ($C_{\rm grp}$) of main- or inter-pulses of MP-GRP or IP-GRPs with those of corresponding NORMAL cycles ($C_{\rm nor}$). 
In this comparison, we defined four types of phase widths ($\Delta \varphi$) to accumulate the peak counts; i.e., $\Delta \varphi = $ 0.20 phases (covering main- or inter pulses), 1/11, 1/31, and 1/128 phases.
The enhancement of $C_{\rm grp}$ from $C_{\rm nor}$ accumulated within $\Delta \varphi$ can be defined as $\xi(\Delta \varphi) \equiv \frac{C_{\rm grp}(\Delta \varphi) - C_{\rm nor}(\Delta \varphi)}{C_{\rm nor}(\Delta \varphi)}$. 
The table~\ref{table:grp_enhancement_stat} summarize $\xi(\Delta \varphi)$ of each $\Delta \varphi $, shown in the percentage, for each instrument, with the significance to the statistical errors. 
As a result, no larger than a 2 sigma enhancement was detected around GRPs in all cases.
The fluctuation got smaller when we restricted the phase width for MP-GRPs 
due to the sharp pulse profile of the main pulse,
except for the $\Delta \varphi = 1/128 \sim 0.008$ phase-width cases with poorer photon statistics,
although such a trend could not be seen for the inter-pulses that had a shallower shape.

\begin{table*}
  \tbl{Summary of enhancements in X-ray flux at GRPs}{%
  \begin{tabular}{llcccc}
      \hline
      Instrument & pulse &
      0.0078 phase$^\dagger$ & 0.0322 phase$^\dagger$ & 0.091 phase$^\dagger$ & 0.200 phase$^\dagger$ \\
      \hline
      SXS & main pulse   &  1\% (0.0$\sigma$)& 10\% (0.7$\sigma$)&  1\% (0.1$\sigma$)&7\% (0.6$\sigma$)\\
      SXS & inter pulse & 310\% (1.6$\sigma$)& 93\% (1.3$\sigma$)& 55\% (1.1$\sigma$)&31\% (0.6$\sigma$)\\
      HXI & main pulse  & 11\% (0.5$\sigma$)& -2\% (-0.2$\sigma$)& -5\% (-0.6$\sigma$)&6\% (0.7$\sigma$)\\
      HXI & inter pulse & 136\% (1.4$\sigma$)& 78\% (1.3$\sigma$)& 29\% (0.9$\sigma$)&49\% (1.5$\sigma$)\\
      SGD & main pulse  & -116\% (1.2$\sigma$)& -39\% (0.7$\sigma$)& -43\% (-1.0$\sigma$)&-59\% (-1.4$\sigma$)\\
      SGD & inter pulse & 310\% (1.6$\sigma$)& 93\% (1.3$\sigma$) & 55\% (1.1$\sigma$) &31\% (0.6$\sigma$) \\
      Combined & main pulse  & 8.9\% (0.5$\sigma$)& 5.1\% (0.6$\sigma$) & -0.8\% (-0.1$\sigma$) &13.2\% (1.8$\sigma$) \\
      Combined & inter pulse & 195\% (2.2$\sigma$)& 45\% (1.2$\sigma$) & 50\% (1.8$\sigma$) &66.5\% (2.4$\sigma$) \\ 
      \hline
    \end{tabular}}\label{table:grp_enhancement_stat}
\begin{tabnote}
$\dagger$ the phase width $(\Delta \varphi)$.\\
The values represent the enhancement $\xi(\Delta \varphi)$ (\%) and values in parentheses show the significance in the standard deviation of each to the statistical errors. 
\end{tabnote}
\end{table*}

\begin{figure*}
 \begin{center}
  \includegraphics[width=0.75\textwidth,angle=0]{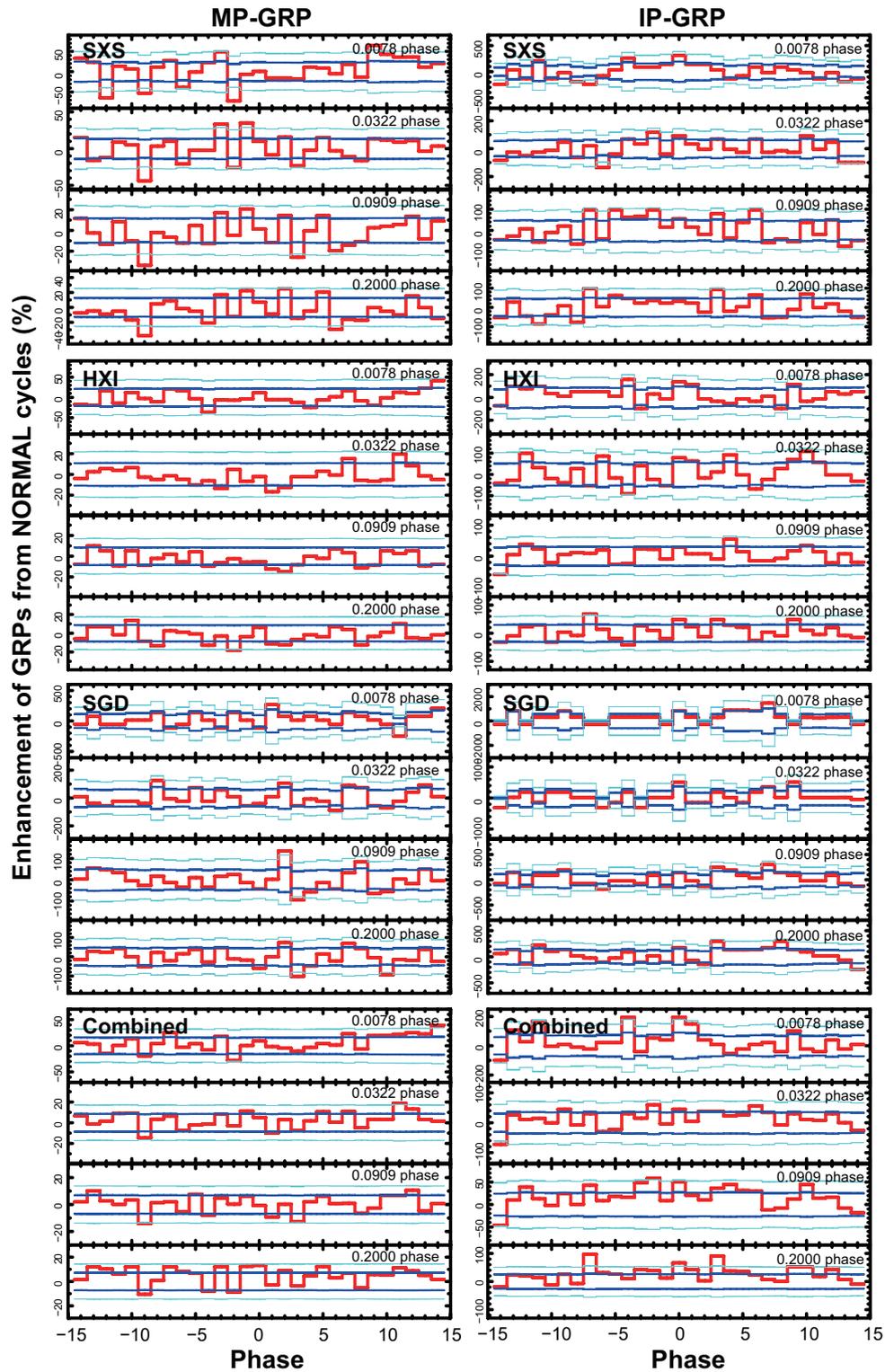}
 \end{center}
\caption{The enhancement of inter or main pulses on the MP-GRP or IP-GRP were shown
in the left and right plots, respectively. 
From the top to bottom panels, the SXS, HXI, SGD-1, and combined data were 
shown, respectively.
The enhancements of GRP relative to normal cycles were measured 
at the corresponding pulse peaks (i.e., inter pulse of main pulse) 
with 0.0078, 0.0322, 0.0909, and 0.200 phase widths, which were shown from top to bottom in each plot.
The enhancement, as a percentage, was shown in red, and statistical uncertainties of 1 and 2 sigma 
were shown in blue and green colors, respectively.}
\label{fig:ana_29pulses_enhance}
\end{figure*}

To test the enhancement $\xi(\Delta\varphi)$ at the snapshot on GRP ($\varphi=0$), 
same trials were repeated for 29 cycles around the GRP, 
i.e. the 14 cycles before to the 14 cycles after the MP-GRP or IP-GRPs 
($-14.5 \leq \varphi \leq 14.5$) as plotted in figure \ref {fig:ana_29pulses_enhance}.
Therefore, a possible enhancement at $\varphi=$ was within the 
fluctuations of $\xi(\Delta \varphi)$ in other cycles 
to within $2 \sigma$ variations for 28+1 cycles.
Numerically, the 3 $\sigma$ upper limits of the variations at the MP-GRP during the main-pulse phases 
(i.e., $\varphi=$ -0.1 -- 0.1, with 0.200 phase-width in figure \ref{fig:ana_29pulses_enhance}) 
will be $\xi_{\rm MPGRP}(0.200{\rm ~phase}) = 40, 30$, and $110$ \% of the X-ray flux in the 
NORMAL cycles, with the SXS, HXI, and SGD, corresponding roughly to the 2 -- 10, 2 -- 80, 10 -- 300 keV bands.
Similarly, the 3 $\sigma$ upper limits for the IP-GRP during inter-pulse phases ($\varphi=$ 0.3 -- 0.5) 
were $\xi_{\rm IPGRP} (0.200 {\rm ~phase}) = 130, 90$, and $420$ \% in the same energy bands listed above, respectively.
When all of the instruments (i.e., the SXS, HXI, and SGD-1) were used for this study, 
the upper-limit values become tighter at $\xi(0.200 {\rm ~phase}) =$ 22\% and 80\% of the NORMAL cycles for the MP- and IP-GRPs, respectively.
In addition, in order to see a possible enhancement on a short-time scale around the peaks of pulses, as had 
been seen in the optical observations \citep{2003Sci...301..493S,2013ApJ...779L..12S}, 
the enhancements of MP- and IP-GRPs accumulated within the $\Delta \varphi = 1/31 \sim$ 0.03 phase-width were also numerically checked, $\xi(0.03 {\rm ~phase}) =$ 25\% and 110\% for MP- and IP-GRPs were obtained.
The 3-$\sigma$ upper limits of $\xi$ from the 29-cycles study were summarized in table~\ref{table:grp_enhancement_summary}.

\begin{table}
  \tbl{Upper limit of enhancement of GRP (3$\sigma$)}{%
  \begin{tabular}{cccc}
      \hline
    Instrument & Energy band & MP-GRP & IP-GRP \\
      \hline
    \multicolumn{4}{l}{\bf $\Delta \varphi =$ 0.2000 phase}\\
    SXS & 2 -- 10 keV   & 40\%  & 130\% \\
    HXI & 5 -- 80 keV   & 30\%  & 90\% \\
    SGD & 10 -- 300 keV & 110\% & 420\%\\
    all & 2 -- 300 keV  & 22\%  & 80\% \\
    \multicolumn{4}{l}{\bf $\Delta \varphi =$ 0.0322 phase}\\
    SXS & 2 -- 10 keV   & 90\%  & 180\% \\
    HXI & 5 -- 80 keV   & 40\%  & 200\% \\
    SGD & 10 -- 300 keV & 200\% & 1100\%\\
    all & 2 -- 300 keV  & 25\%  & 110\% \\
      \hline
    \end{tabular}}\label{table:grp_enhancement_summary}
\begin{tabnote}
The ``all'' represents the sum of SXS, HXI, and SGD instruments.
\end{tabnote}
\end{table}

\subsection{Upper limit of Enhanced Peak flux}
\label{section:analyses_results:upper_limit_flux}
\begin{figure}
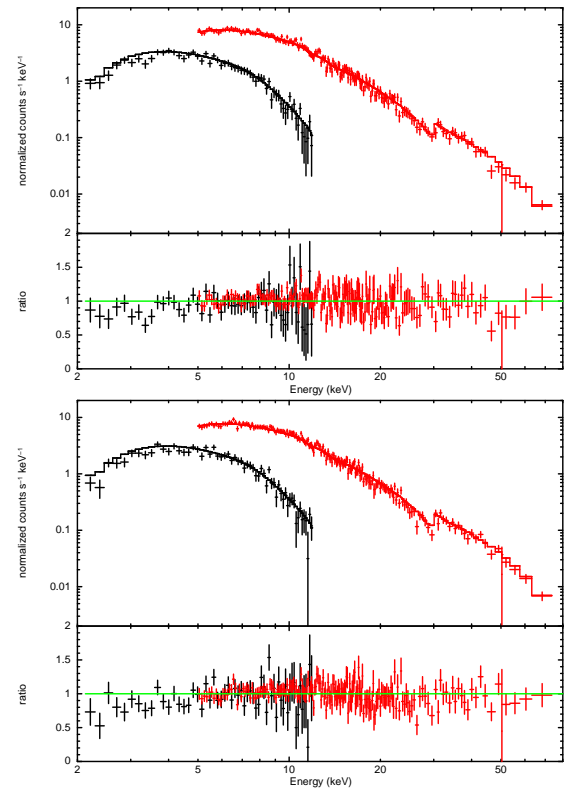

 \begin{center}
  \includegraphics[width=0.30\textwidth,angle=-90]{hitomi_crab_grp_figure06a.eps}
  \includegraphics[width=0.30\textwidth,angle=-90]{hitomi_crab_grp_figure06b.eps}
 \end{center}
\caption{Left and right panels show the X-ray spectra during the main and inter pulses, respectively. 
The spectra with the SXS and the HXI are shown in black and red crosses, respectively.
The best-fit power-law models are shown in red and black lines for the SXS and HXI, respectively.
The bottom panels represent the ratio between the data and the model.}
\label{fig:ana_pulse_spectrum}
\end{figure}

To convert the enhancement of GRP in count rate into an X-ray flux, the X-ray spectra of purely pulsed components (i.e., main and inter pulses) were numerically tested. 
First, the SXS and HXI events were extracted by phases,  $\varphi=$ -0.1 -- 0.1, $\varphi=$ 0.3  E 0.5, and $\varphi=$ 0.6 - E0.8, corresponding to the main-pulse (MP), inter-pulse (IP), and off (OFF) phases, respectively, and the pulse-height distributions were accumulated.
The dead time correction was applied to the HXI data with the Hitomi ftools, {\it hxisgddtime}; the live time of the HXI-1 and HXI-2 were 73.9 \% and 76.6 \% for this observation.
Only the high-primary and the medium-primary grades (Hp and Mp grades, respectively, defined in \cite{ASTROH_SXS}) were accumulated in the SXS spectral analyses here in order to reduce systematic errors in the response matrix. 
The X-ray spectra of the pure pulsed components were calculated by subtraction of the OFF-phase spectrum from the MP or IP spectra.
Thanks to the fine timing resolutions of the SXS, HXI, and SGD \citep{2017Hitomi_Time_system}, the X-ray spectra of the pure-pulsed components were clearly demonstrated in figure \ref{fig:ana_pulse_spectrum}.

\begin{table}
  \tbl{Pulsed flux of Crab pulsar}{%
  \begin{tabular}{ccccc}
      \hline
    Instrument & Energy band & $\Delta\varphi$ & Main pulse$^\dagger$ & Inter pulse$^\dagger$\\
      \hline
   SXS& 2 -- 10 keV   & 0.20 & $31 \pm 5$   & $30 \pm 4$ \\
   HXI& 5 -- 80 keV   & 0.20 & $59_{-8}^{+9}$  & $62 \pm 9$ \\
   SGD& 10 -- 300 keV & 0.20 & $76_{-10}^{+11}$ & $87 \pm 13$ \\
   all& 2 -- 300 keV  & 0.20 & $108_{-15}^{+16}$& $116 \pm 17$\\
      \hline
   SXS& 2 -- 10 keV   & 0.03 & $4.0 \pm 0.6$& $2.7 \pm 0.4$\\
   HXI& 5 -- 80 keV   & 0.03 & $7.0 \pm 1.0$& $5.0 \pm 0.8$ \\
   SGD& 10 -- 300 keV & 0.03 & $10 \pm 2$   & $6.9 \pm 1.5$ \\
   all& 2 -- 300 keV  & 0.03 & $13 \pm 2$   & $9.0 \pm 1.5$ \\
      \hline
    \end{tabular}}\label{table:pulsed_flux_summary}
\begin{tabnote}
$\dagger$ X-ray flux in $10^{-12}$ erg cm$^{-2}$ at the energy band accumulated within the phase-with ($\Delta\varphi$).
\end{tabnote}
\end{table}

To perform spectral fitting of the MP and IP spectra, the spectral response matrices were generated with the Hitomi ftools {\it sxsmkrmf} and {\it aharfgen}, with the exposure map calculated for the HXI and SXS using the ftools {\it sxsregext} and {\it ahexpmap}, respectively. 
The result was that the MP and IP spectra were well reproduced by a single power-law model with a photon indices of $1.94 \pm 0.02$ and $ 1.87 \pm 0.02$ and X-ray fluxes of $(4.7 \pm 0.1) \times 10^{-9}$ and $(4.4 \pm 0.5) \times 10^{-9}$ ergs cm$^{-2}$ s$^{-1}$ in the 2 -- 10 keV band with the reduced $\chi^2$ of 305.85 and 301.18 for 250 degrees of freedom, respectively, as shown in the figure \ref{fig:ana_pulse_spectrum}.
The pulsed flux obtained with Hitomi accumulated within the time interval of $\Delta\varphi$ phase was summarized in table~\ref{table:pulsed_flux_summary}.
Therefore, the 3 $\sigma$ upper limit values of enhancement in terms of flux
can be obtained by multiplying the values in table~\ref{table:grp_enhancement_summary} (section \ref{section:analyses_results:pulse_enhancement}) with those in table~\ref{table:pulsed_flux_summary}. The upper limits of enhancements of MP-GRPs in the X-ray flux in the 2 -- 300 keV band within the phases of 0.20 or 0.03 become $(24$ or $3.3) \times 10^{-12}$ erg cm$^{-2}$, respectively, and the same for IP-GRPs were $(93$ or $9.9) \times 10^{-12}$ erg cm$^{-2}$, respectively.

\section{Discussion}
\label{section:discussion}

\begin{figure}
 \begin{center}
  \includegraphics[width=0.45\textwidth]{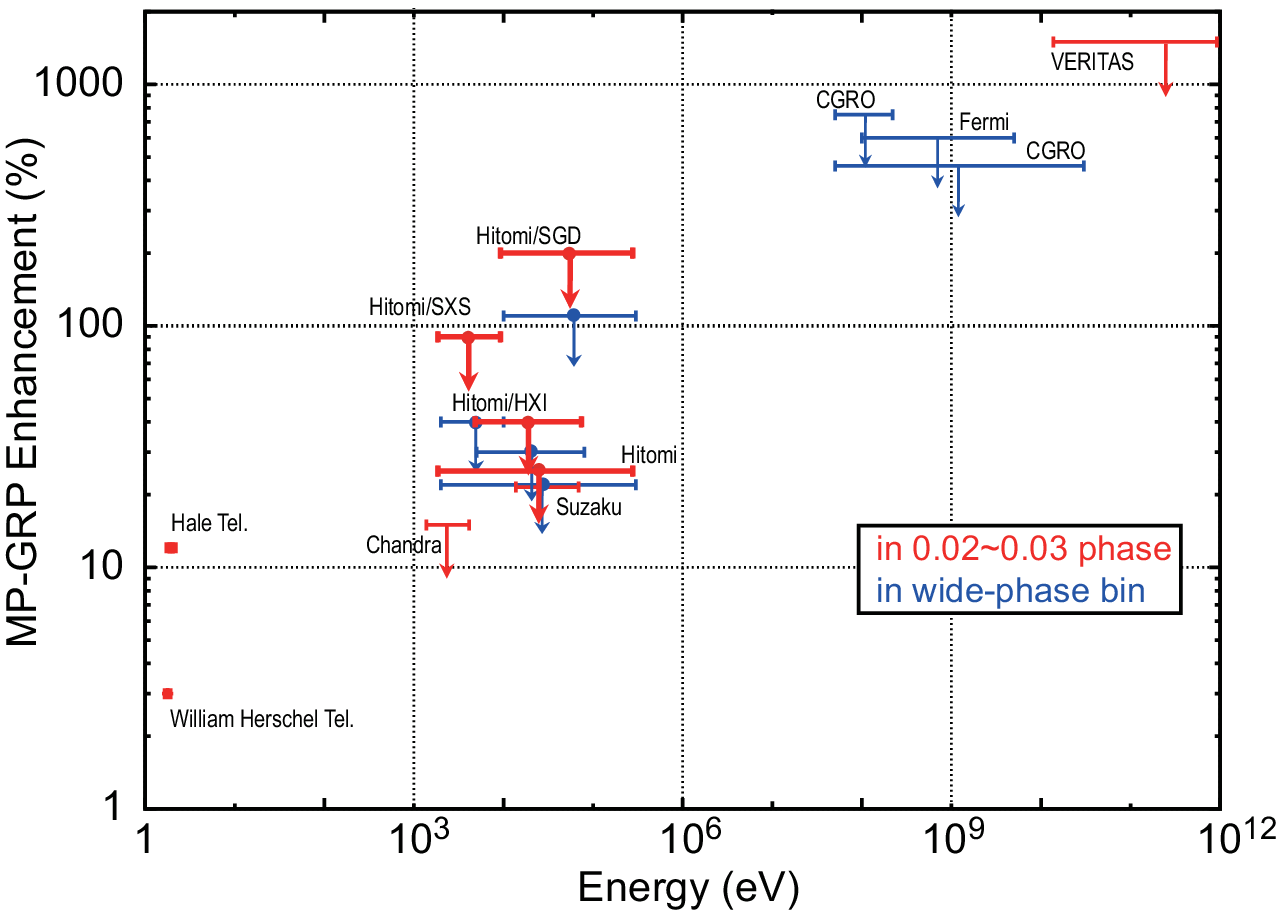}
 \end{center}
\caption{The enhancement of MP-GRPs of Crab pulsar in various energy bands obtained by the 
Hilliam Herschel Telescope \citep{2003Sci...301..493S}, 
Hale telescope \citep{2013ApJ...779L..12S},
Chandra \citep{2012ApJ...749...24B},
Suzaku \citep{2014JPSCP.1.015106},
Hitomi (this work),
CGRO \citep{1995ApJ...453..433L,1998ApJ...496..863R},
Fermi \citep{2011ApJ...728..110B}, and VERITAS \citep{2012ApJ...760..136A}
All the upper limits shown in arrows represent 3 $\sigma$ values.
The red and blue cases indicate the enhancement measured in a short-phase width ($\Delta\varphi = $ 0.02 -- 0.03 phase) and wider-phase width ($\Delta\varphi > 0.1$), respectively.
Note that the thresholds of GRP detections in the radio band were different among them.}
\label{fig:grp_enhancement_summary}
\end{figure}

With the simultaneous observations of the Crab with Hitomi and Kashima radio observatory, the correlation studies in the X-ray band with about 1,000 GRPs have been performed (Section \ref{section:observation_dataReduction}). 
No significant changes in the X-ray pulse profiles were detected along all the phase bins at the GRP cycles (Section \ref{section:analyses_results}), and the 3-$\sigma$ upper-limit values for the MP-GRPs in the 2 -- 300 keV band with Hitomi were $\xi =$ 22\% and 80\% within the time interval of $\Delta\varphi = 0.200$ phase, as summarized in table~\ref{table:grp_enhancement_summary}.
The upper limits in the 4.5 -- 10 keV and the 70 -- 300 keV were obtained for the first time with Hitomi, and those in other bands were consistent with previous works \citep{2014JPSCP.1.015106,2012ApJ...749...24B} as shown in Fig.\  \ref{fig:grp_enhancement_summary}.
Our result constitutes the second case of a study in the hard X-ray band 
where the flux (in $\nu F_\nu$ space) of the pulsed component of the Crab pulsar became the highest, 
following a previous report of a marginal detection at the $2.7 \sigma$ level with Suzaku \citet{2014JPSCP.1.015106}.
Our results were mainly limited by the photon numbers in the X-ray band, 
and the statistical errors dominated the results.
The pulse shape in the X-ray band was observationally confirmed to be stable with the 1 $\sigma$ fluctuation of $\sim 0.7$ \% level by RXTE showing about-two-times intensive pulses \citep{1999ApJ...522..440P,2001A&A...373..236V}, which could be shallow ``giant X-ray pulses'' (GXPs) but the timing correlations between these shallow GXPs and the GRPs were unknown. 
In our X-ray correlation study using the Hitomi satellite, we could not identify such GXPs due to a poor effective area.

As described in section \ref{section:analyses_results:pulse_enhancement},
no significant variabilities were detected in the X-ray pulse profiles of
14 cycles before and after the GRPs.
Similarly, in the optical band, 
the enhancements related to the GRPs only happen 
in narrow time intervals ($\sim 100~\mu$s) at the pulse peak, 
and the pulse profiles in other phases were stable \citep{2003Sci...301..493S}.
These facts indicated that the magnetosphere is stable during the GRPs, which should originate from a local place within the magnetosphere. 

What happens during the GRP on the pulsar,
when the structure of the magnetosphere does not change?
Here, we assumed the emission mechanism of the optical pulses is 
synchrotron emission, like X-ray pulses, 
because the optical emission seems to have the same origin 
as that of the X-rays from the multi-wavelength spectrum of 
the pulsed component of the pulsar (e.g. see \cite{2014RPPh...77f6901B}). 
To increase the synchrotron emission temporarily on a short time scale 
of $\mu$s, whilst maintaining the structure of the magnetosphere, 
only two candidates can be considered; 
a) an increase in the number of particles for radiation, or 
b) a change in the local magnetic field strength.
However, case b) would be considered difficult to achieve normally,
and the pulse phase needs not be aligned to the main or inter pulses, 
and therefore it is straightforward to think that 
the case a) is the origin of the optical enhancement at GRPs.
Such occasion might occur after a magnetic reconnection near the light cylinder \citep{2004IAUS..218..369I} resulting a higher density plasma than the Goldreich-Julian density in the GRP region \citep{2007MNRAS.381.1190L}.
In the Crab pulsar, the emission regions for the radio, optical, and X-rays are 
normally considered to be close to each other 
because the pulses are well aligned in these energy bands, 
although the pulse profile in the X-ray band is wider than that in radio.
If the number of particles for synchrotron emission 
increases in a local region that emits very short GRPs, 
a possible X-ray enhancement should also occur very shortly,
within about 10 $\mu$s just on the pulse peak, 
like the optical and radio cases.
If such short enhancement will be detected in the X-ray band,
we are able to reinforce the idea a), 
but the fine pulse profiles divided by 1/128 phases with Hitomi 
(Fig.\ \ref{fig:ana_1pulse}) do not show such enhancement at the peak statistically.

Finally, we discuss the energy balance between the X-ray upper limit of the pulse-peak flux ($\xi$) and the radiation energy of GRP in the radio band, $E_{\rm radio}$.
As described in section \ref{section:analyses_results:pulse_peak_enhancement}, the enhancement at the pulse peak $\xi$ were within the 2 $\sigma$ fluctuation among the 29 cycles around GRPs, and the upper limit of $\xi$ in flux were obtained at about $3.3 \times 10^{-12}$ erg cm$^{-2}$ accumulated in the time interval of $\Delta\varphi = 0.03$ phase.
The threshold of detection of GRPs in our radio observation was 2.2 kJy$\cdot \mu$s 
(Section \ref{section:GRPselection}), 
which corresponds to a total emission-energy par area of 
$E_{\rm radio} > 2.2 \times 10^{-17}$ erg cm$^{-2}$ 
in a 10 $\mu$s accumulation under the assumption that the GRP pulses emit 
with 1 GHz-width in the radio \citep{2016ApJ...832..212M}.
Interestingly, the optical enhancement of $\xi_{\rm opt}(0.003 {\rm ~phase}) = $ 3\% reported by \citet{2003Sci...301..493S} within 100~$\mu$s bin corresponds to a roughly-equivalent energy with $E_{\rm radio}$.
Therefore, here we assume that the radiation energy of possible X-ray enhancement is almost equivalent to $E_{\rm radio}$, although the pulsed energy spectrum of Crab (e.g. \cite{2014RPPh...77f6901B}) indicates that the optical light and X-rays have different origins.
If an X-ray detection is performed with rather wide phase-bins ($\Delta\varphi = $ 0.20 phase around the peak), it appears at $\xi_{\rm X}(0.20 \rm{~phase}) = 2 \times 10^{-5}$~\% of the X-ray normal pulses in the 2 -- 300 keV band within the same phase-bins. 
This enhancement appears better at the $\xi_{\rm X} (0.0003 \rm{~phase}) = 0.02$~\% of normal pulse flux when the X-ray observation can resolve the 10 $\mu$s time-bin at the pulse peak.
But the value is still undetectable under the poor statistics of our Hitomi data and the timing accuracy \citep{2017Hitomi_Time_system}.
Therefore, the results did not statistically rule out variations correlated with the GRPs, 
because the possible X-ray enhancement may appear as $>0.02$\% brightening of the pulse peak under such conditions. 
We can expect future X-ray missions with larger effective area 
and better timing capability, 
such as the recently-launched NICER mission \citep{NICER}, 
for continuing the X-ray correlation studies of GRPs.
If the GXPs appears in short-phase bins and are correlated with GRPs,
NICER may detect the enhancement in X-ray band,
althoufh the sensitivity peak of NICER is at a softer energy band than 
that of the previous, larger area mission RXTE, so the count rate expected for Crab pulses is comparable.

\section*{Acknowledgements}
We thank the support from the JSPS Core-to-Core Program.
We acknowledge all the JAXA members who have contributed to the ASTRO-H (Hitomi)
project.
All U.S. members gratefully acknowledge support through the NASA Science Mission
Directorate. Stanford and SLAC members acknowledge support via DoE contract to SLAC
National Accelerator Laboratory DE-AC3-76SF00515. Part of this work was performed under
the auspices of the U.S. DoE by LLNL under Contract DE-AC52-07NA27344.
Support from the European Space Agency is gratefully acknowledged.
French members acknowledge support from CNES, the Centre National d'\'{E}tudes Spatiales.
SRON is supported by NWO, the Netherlands Organization for Scientific Research.  Swiss
team acknowledges support of the Swiss Secretariat for Education, Research and
Innovation (SERI).
The Canadian Space Agency is acknowledged for the support of Canadian members.  
We acknowledge support from JSPS/MEXT KAKENHI grant numbers 15J02737,
15H00773, 15H00785, 15H02090, 15H03639, 
15K05088, 
15K05069, 
15H05438, 15K05107, 15K17610,
15K17657, 16J00548, 16J02333, 
16J06773, 
16H00949, 16H06342, 16K05295, 16K05296,
16K05300, 16K13787, 16K17672, 16K17673, 21659292, 23340055, 23340071,
23540280, 24105007, 24244014, 24540232, 25105516, 25109004, 25247028,
25287042, 25400236, 25800119, 26109506, 26220703, 26400228, 26610047,
26800102, JP15H02070, JP15H03641, JP15H03642, JP15H06896,
JP16H03983, JP16K05296, JP16K05309, JP16K17667, and JP16K05296.
The following NASA grants are acknowledged: NNX15AC76G, NNX15AE16G, NNX15AK71G,
NNX15AU54G, NNX15AW94G, and NNG15PP48P to Eureka Scientific.
H. Akamatsu acknowledges support of NWO via Veni grant.  
C. Done acknowledges STFC funding under grant ST/L00075X/1.  
A. Fabian and C. Pinto acknowledge ERC Advanced Grant 340442.
P. Gandhi acknowledges JAXA International Top Young Fellowship and UK Science and
Technology Funding Council (STFC) grant ST/J003697/2. 
Y. Ichinohe, K. Nobukawa, and H. Seta are supported by the Research Fellow of JSPS for Young
Scientists.
N. Kawai is supported by the Grant-in-Aid for Scientific Research on Innovative Areas
``New Developments in Astrophysics Through Multi-Messenger Observations of Gravitational
Wave Sources''.
S. Kitamoto is partially supported by the MEXT Supported Program for the Strategic
Research Foundation at Private Universities, 2014-2018.
B. McNamara and S. Safi-Harb acknowledge support from NSERC.
T. Dotani, T. Takahashi, T. Tamagawa, M. Tsujimoto and Y. Uchiyama acknowledge support
from the Grant-in-Aid for Scientific Research on Innovative Areas ``Nuclear Matter in
Neutron Stars Investigated by Experiments and Astronomical Observations''.
N. Werner is supported by the Lend\"ulet LP2016-11 grant from the Hungarian Academy of
Sciences.
D. Wilkins is supported by NASA through Einstein Fellowship grant number PF6-170160,
awarded by the Chandra X-ray Center, operated by the Smithsonian Astrophysical
Observatory for NASA under contract NAS8-03060.

We thank contributions by many companies, including in particular, NEC, Mitsubishi Heavy
Industries, Sumitomo Heavy Industries, and Japan Aviation Electronics Industry. Finally,
we acknowledge strong support from the following engineers.  JAXA/ISAS: Chris Baluta,
Nobutaka Bando, Atsushi Harayama, Kazuyuki Hirose, Kosei Ishimura, Naoko Iwata, Taro
Kawano, Shigeo Kawasaki, Kenji Minesugi, Chikara Natsukari, Hiroyuki Ogawa, Mina Ogawa,
Masayuki Ohta, Tsuyoshi Okazaki, Shin-ichiro Sakai, Yasuko Shibano, Maki Shida, Takanobu
Shimada, Atsushi Wada, Takahiro Yamada; JAXA/TKSC: Atsushi Okamoto, Yoichi Sato, Keisuke
Shinozaki, Hiroyuki Sugita; Chubu U: Yoshiharu Namba; Ehime U: Keiji Ogi; Kochi U of
Technology: Tatsuro Kosaka; Miyazaki U: Yusuke Nishioka; Nagoya U: Housei Nagano;
NASA/GSFC: Thomas Bialas, Kevin Boyce, Edgar Canavan, Michael DiPirro, Mark Kimball,
Candace Masters, Daniel Mcguinness, Joseph Miko, Theodore Muench, James Pontius, Peter
Shirron, Cynthia Simmons, Gary Sneiderman, Tomomi Watanabe; ADNET Systems: Michael
Witthoeft, Kristin Rutkowski, Robert S. Hill, Joseph Eggen; Wyle Information Systems:
Andrew Sargent, Michael Dutka; Noqsi Aerospace Ltd: John Doty; Stanford U/KIPAC: Makoto
Asai, Kirk Gilmore; ESA (Netherlands): Chris Jewell; SRON: Daniel Haas, Martin Frericks,
Philippe Laubert, Paul Lowes; U of Geneva: Philipp Azzarello; CSA: Alex Koujelev, Franco
Moroso.

\begin{contribution}
Y. Terada led this study in data analysis and writing drafts,
in addition to the {\it Hitomi} timing Calibration and software preparation.
X-ray data analyses and calibrations were done with 
T. Enoto, S. Koyama, A. Bamba S. Nakashima, T. Yaqoob, 
H. Takahashi, S. Watanabe, and K. Oshimizu.
T. Terasawa led the radio data analysis
with M. Sekido, K. Takehuji, E. Kawai, H. Misawa,
F. Tsuchiya, R. Yamazaki, E. Kobayashi, S. Kisaka, and T. Aoki.
T. Dotani, L. Gallo, R. Mushotzky, C. Ferrigno, K. Pottschmidt, M. Loewenstein, 
M. Tsujimoto, and S. Safi-Harb improved the draft.
\end{contribution}


\appendix 
\section{Details for radio data reduction}
\label{section:appendix_radio}

\subsection{TEMPO2 timing package}
\label{section:TEMPOtwo}
The pulsar timing package TEMPO2 (Hobbs et al., 2006)
gives the function ${\tilde f}(t_{\rm UTC})$ to convert the observation time in UTC, $t_{\rm UTC}$, to 
the time at the solar system barycenter (TDB), ${\tilde t}_{\rm TDB}$.
TEMPO2 also gives the signal frequency ${\nu}_{\rm ISM}$ 
in the rest frame of the interstellar matter (ISM)
corresponding to the observation frequency $\nu_{\rm obs}$,
which is Doppler shifted by the revolutionary+rotationary motion of the earth with respect to the solar system barycenter.
We defined the Doppler factor $\eta  \equiv \nu_{\rm obs}/{\nu}_{\rm ISM}$.

\subsection{Calculations of S/N}
We dedispersed the raw antenna voltage data of the channel $k$, $V^{\rm raw}_k(t)$, 
to obtain $V_k(t)$. 
We integrated\footnote{
For simplicity we use the term, 'integrate'.
In reality, of course, we calculated equation (\ref{eqn:VoltageSquared})
as finite sums over $t_n \le t <t_{n+1}$ with an original sampling time step $\delta t$ =15.625ns.
} 
$|V_k(t)|^2$ over the time interval of $\Delta t$=10 $\mu$s,
\begin{equation}
{\cal E}_k (t_n) = \frac{1}{\Delta t} \int_{t_n}^{t_{n+1}} |V_k(t')|^2 dt'
\label{eqn:VoltageSquared}
\end{equation}
where we defined
the binned times as $t_n=t_{\rm start} + n \Delta t$
(Here we represent $t$ and $t_n$ by UTC.
$t_{\rm start}$=12:15:00 UT.)
Between the channel $k$ and  6,
there is an arrival time difference owing to the propagation group delay,
\begin{equation}
\tau_{k,6} = \frac{e^2}{2\pi m_e c} DM
  \left \{
    \frac{1}{\nu_k^2} - \frac{1}{\nu_{6}^2}
  \right \}
  \eta
\label{eqn:GroupDelay}
\end{equation}
where ($e$, $m_e$, $c$) are usual physical quantities, $DM$ the dispersion measure,
and $\nu_k$ and $\nu_{6}$ the highest frequencies of the bands $k$ and $6$.
The Doppler factor $\eta$ 
appears in equation (\ref{eqn:GroupDelay})
since $\tau_{k,6}$, $\nu_k$, and $\nu_6$ are defined in the observer's frame.
We combined incoherently
${\cal E}_0 (t_n)$, ${\cal E}_1 (t_n)$,..,
${\cal E}_6 (t_n)$,
as,
\begin{equation}
{\cal E}_{\rm sum} (t_n) = \sum_{k=0,1,4,5,6} {\cal E}_{k} (t_n-\tau_{k,6})
\label{eqn:incoheretsum}
\end{equation}
where appropriate interpolations were taken to calculate the RHS of equation (\ref{eqn:incoheretsum}).

We calculated the average $\bar {\cal E}_{\rm sum}$ and standard deviation $\sigma_{\rm sum}$
of ${\cal E}_{\rm sum}(t_n)$ over an appropriate longer time interval ($\Delta T$, for which we take 1s).
Ideally $\bar {\cal E}_{\rm sum}$ and $\sigma_{\rm sum}$ are constant in time.
In reality, however, they showed slight and gradual variations in time $T_M (= t_{\rm start} + M \Delta T; M=0,1,2...)$.
We calculate the signal-to-noise ratio (S/N) at $t_n$ for $T_M \le t_n < T_{M+1}$ as 
\begin{equation}
{\cal S}_{\rm sum}(t_n) = \frac{1}{\sigma_{\rm sum}(T_M)} ({\cal E}_{\rm sum}(t_n) - {\bar{\cal E}_{\rm sum}}(T_M))
\label{eqn:SNdefinition}
\end{equation}
With a given threshold $\cal{S}_{\rm sum, \rm thr}$,
we selected `GRP candidates' for each one in which there was an enhancement\footnote{
A strong GRP gives enhancements of the average and standard deviation,
so that the S/N obtained by equation (\ref{eqn:SNdefinition}) is reduced.
To avoid this effect, we replace 
the average and standard deviation in equation (\ref{eqn:SNdefinition})
with the values interpolated from those
obtained in the surrounding time intervals that are unaffected by GRPs.
},
${\cal{S}}_{\rm sum}(t_n) >= \cal{S}_{\rm sum, \rm thr}$.

\subsection{Time-domain RFI rejection}
To eliminate RFI further,
we conducted the following auxiliary process:
We calculated the squared antenna voltages ${\cal E}^{\rm raw}_k (t_n)$ 
following equation (\ref{eqn:VoltageSquared})
except that we used $V^{\rm raw}_k(t)$ instead of $V_k(t)$.
From ${\cal E}^{\rm raw}_k (t_n)$ we then calculated
\begin{equation}
{\cal E}^{\rm raw}_{\rm sum} (t_n) \equiv  \sum_{k=0,1,4,5,6} {\cal E}^{\rm raw}_k (t_n)
\end{equation}
and their average and standard deviation, ${\bar {\cal E}}^{\rm raw}_{\rm sum}(T_M)$ and $\sigma^{\rm raw}_{\rm sum}(T_M)$.
We watched the behaviors of $\sigma_{\rm sum}(T_M)$ and $\sigma^{\rm raw}_{\rm sum}(T_M)$
throughout the observation interval.
When $\sigma_{\rm sum}(T_M) < \sigma^{\rm raw}_{\rm sum}(T_M)$,
we reject the data 
${\cal{S}}_{\rm sum}(t_n)$ for $T_M \le t_n <T_{M+1}$, as affected by RFIs.

\subsection{Rotation phase and GRP identification}
We calculated the phase $\varphi_n$ of a GRP candidate at the time in UTC $t_n$, 
\begin{equation}
\varphi_n = frac(y)
~~~{\rm with}~~~
y         = y_0 + \nu_{\rm rot} {\tilde t}_n + 0.5 {\dot \nu} _{\rm rot} {\tilde t}_n^2,
\label{eqn:getPhi}
\end{equation}
where 
$\tilde t_n = {\tilde f}(t_n)$ is the time in TDB,
$frac(y)$ the fractional part of $y$,
$y_0$ the initial phase at 00:00:00 TDB, 
$\nu_{\rm rot}$ and ${\dot \nu}_{\rm rot}$ the rotation frequency and its time derivative from 
the Jodrell bank monthly ephemeris (table~\ref{table:FreqTable}).
In the operation of equation (\ref{eqn:getPhi}), 
we also recorded the integer part of $y$ as the sequential pulse number of the day, $N_{\rm pulse}$,
which is to be used for the GRP and X-ray photon comparison.

As discussed in section~\ref{section:GRPselection}
we classify the GRP candidates according to their values of $\varphi$
by setting two selection ranges, 
($\varphi_{\rm MP,1}$, $\varphi_{\rm MP,2}$) for main pulse GRPs, and
($\varphi_{\rm IP,1}$, $\varphi_{\rm IP,2}$) for interpulse GRPs.
With the choice of $\Delta t$=10$\mu$s,
two thirds of GPR candidates in the 1.4-1.7GHz band are isolated in the $\varphi$ space.
However, for the remaining one third of (stronger) GRPs,
2$\sim$4 GRP candidates of the same $N_{\rm pulse}$
are found in the same selection range for $\varphi$.
For such cases, we count them as one GRP, either main pulse or
interpulse.

For a GRP occurring near the binned time boundary $t=t_n$,
its contribution is divided into ${\cal E}_{\rm sum} (t_n)$ and ${\cal E}_{\rm sum} (t_{n+1})$,
and the corresponding ${\cal S}_{\rm sum}(t_n)$ and ${\cal S}_{\rm sum}(t_{n+1})$
are artificially lowered (sometimes both are less than ${\cal S}_{\rm sum ,thr}$).
To avoid mal-counting of GRPs caused by this effect,
we repeated the procedure from equation (\ref{eqn:VoltageSquared}) to equation (\ref{eqn:SNdefinition})
for the binned time with a shift of $\Delta t/2$:
Firstly, we calculate
$$
~~~~~~~~~~~~~~
{\cal E}_k (t_{n+\frac{1}{2}}) = \frac{1}{\Delta t} 
  \int_{t_n+\frac{\Delta t}{2}}^{t_{n+1}+\frac{\Delta t}{2}} |V_k(t')|^2 dt'
~~~~~~~~~~~~~~~~~~~~~~~~~~~~~~~~~~~~~~~~~~~~~~~~~~~~~~~~~
~~~~~~~~~~~~~~
~~~~~~
{\rm equation (\ref{eqn:VoltageSquared}')}
$$
If the resultant ${\cal S}_{\rm sum}(t_{n+\frac{1}{2}})$ exceeds ${\cal S}_{\rm sum, thr}$,
this GRP is `rescued from the sea of the noise'.
We found that about 10\% of GRPs are thus rescued.

\subsection{Radiometer equation}
The flux density ${\cal F}$ is calculated as
$C \times (S/N)$
with $C$ given by the radiometer equation (Dicke, 1946; Lorimer and Kramer, 2004),
\begin{equation}
C=\frac{\mathrm{SEFD}+S_{\rm CN}} {\sqrt{\Delta \nu_{\rm sum} \Delta t}}~~~~{\rm [Jy]}
\label{eqn:radiometer}
\end{equation}
where $\mathrm{SEFD}$ is the system equivalent flux density, and
$S_{{\rm CN}}$ is the flux density of the Crab nebula.
With the representative values,
$\mathrm{SEFD}= 500$Jy and $S_{{\rm CN}}= 810$Jy (Mikami et al., 2016),
we get $C = 40.2$Jy
for $\Delta \nu_{\rm sum}$=106.36MHz (section~\ref{section:FrequencyDomainRFIrejection})
and $\Delta t$=10$\mu$s.
The GRP threshold $S/N=5.5$ used in section~\ref{section:RadioObservation}
corresponds to a flux density threshold of ${\cal F}= 220$Jy,
or a pulse energy threshold ${\cal F}\Delta t=2.2$kJy~$\mu$s.

\section{Extraction of the SGD photo absorption mode}
\label{section:appendix:sgd}

On the timing analyses of the SGD-1 data 
in section \ref{section:observation_dataReduction:data_xray}, 
the photo-absorption events were extracted 
from the unscreened event files to have more effective areas than those of the 
standard Compton scattered events; 
an expression that "FLAG\_LCHKMIO==b0 \&\& FLAG\_CCBUSY[1]==b0 \&\& FLAG\_CCBUSY[2]==b0 \&\& FLAG\_CCBUSY[3]==b0 \&\& FLAG\_HITPAT[1]==b0 \&\& FLAG\_HITPAT[2]==b0 \&\& FLAG\_HITPAT[3]==b0 \&\& FLAG\_HITPAT[4]==b0 \&\& FLAG\_FASTBGO[1]==b0 \&\& FLAG\_FASTBGO[2]==b0 \&\& FLAG\_FASTBGO[3]==b0 \&\& FLAG\_FASTBGO[4]==b0 \&\& FLAG\_SEU==b0 \&\& FLAG\_LCHK==b0 \&\& FLAG\_CALMODE==b0 \&\& FLAG\_TRIGPAT[29]==b0 \&\& CATEGORY==85 \&\& MATTYPE==1 \&\& NUMSIGNAL==1 " were applied to the ufa event and standard GTI in the 2nd extension of the cleaned events were also applied. 


\end{document}